\documentclass[11pt]{article}
\usepackage{amsmath,amssymb,euscript,mathrsfs}
\usepackage{graphicx}

\usepackage{amssymb}
\usepackage{graphicx,hyperref}
\usepackage{amsfonts,bbm}
\usepackage{amsmath,subfigure}
\usepackage[nospace]{cite}                      

\usepackage[top=1in,bottom=1in,left=1in,right=1in]{geometry}

\newcommand{\sub}[2]{#1_{#2}}

\def\obs{x}
\def\revObs{\obs^{(r)}}
\def\subObs#1{\sub{\obs}{#1}}
\def\subRevObs#1{\sub{\revObs}{#1}}
\def\state{S}
\def\model{\mathcal{M}}
\def\cH{\mathcal{H}}
\def\bbP{\mathbb{P}}

\def\bbE{\mathbb{E}}
\def\bbOne{\mathbbm{1}}

\def\recom{\mathcal{R}}

\newcommand{\coal}[1]{C_{#1}}
\newcommand{\float}[1]{C_{> #1}}
\newcommand{\fgreaterthan}[2]{f\left(\subObs{1: #2},R_{\leq #1},C_{> #1}\right)}
\newcommand{\fsum}[2]{f\left(\subObs{1: #2}, T_{#2} > #1\right)}
\newcommand{\feq}[2]{f\left(\subObs{1: #2}, T_{#2} = #1\right)}

\begin{document}

\begin{center}
{\Large \bf Decoding coalescent hidden Markov models in linear time}

\vspace{6mm}
Kelley Harris$^1$, Sara Sheehan$^2$, John A. Kamm$^3$, and Yun S. Song$^{2,3,4}$

\vspace{5mm}
$^1$Department of Mathematics,  University of California, Berkeley\\
$^2$Computer Science Division,  University of California, Berkeley\\
$^3$Department of Statistics,  University of California, Berkeley\\
$^4$Department of Integrative Biology, University of California, Berkeley\\

\vspace{3mm}
{kharris@math.berkeley.edu, ssheehan@eecs.berkeley.edu, jkamm@stat.berkeley.edu, yss@eecs.berkeley.edu}
\end{center}

\vspace{1mm}
\begin{abstract} 
In many areas of computational biology, hidden Markov models (HMMs) have been used to model local genomic features.
In particular, coalescent HMMs have been used to infer ancient population sizes, migration rates, divergence times, and other parameters such as mutation and recombination rates.
As more loci, sequences, and hidden states are added to the model, however, the runtime of coalescent HMMs can quickly become prohibitive.  Here we present a new algorithm for reducing the runtime of coalescent HMMs from quadratic in the number of hidden time states to linear, without making any additional approximations.
Our algorithm can be incorporated into various coalescent HMMs, including the popular method PSMC for inferring variable effective population sizes.
Here we implement this algorithm to speed up our demographic inference method diCal, which is equivalent to PSMC when applied to a sample of two haplotypes. We demonstrate that the linear-time method can reconstruct a population size change history more accurately than the quadratic-time method, given similar computation resources. We also apply the method to data from the 1000 Genomes project, inferring a high-resolution history of size changes in the European population.

\end{abstract}

\vspace{5mm}
{To appear in the Proceedings of the 18th Annual International Conference on Research in Computational Molecular Biology (RECOMB 2014).
The final publication is available at \url{link.springer.com}.}

\section{Introduction}
The hidden Markov model (HMM) is a natural and powerful device for learning functional and evolutionary attributes of DNA sequence data. Given an emitted sequence of base pairs or amino acids, the HMM is well-suited to locating hidden features of interest such as genes and promotor regions \cite{genscan,chromHMM}. HMMs can also be used to infer hidden attributes of a collection of related DNA sequences. In this case, emitted states are a tuple of A's, C's, G's and T's, and the diversity of emitted states in a particular region can be used to infer the local evolutionary history of the sequences. When two sequences are identical throughout a long genetic region, they most likely inherited that region identical by descent from a recent common ancestor. Conversely, high genetic divergence indicates that the sequences diverged from a very ancient common ancestor \cite{beagleIBD,PSMC}. 

In recent years, coalescent HMMs such as the Pairwise Sequentially Markov Coalescent (PSMC)  \cite{PSMC} have been used to infer the sequence of times to most recent common ancestor (TMRCAs) along a pair of homologous DNA sequences. Two other coalescent HMMs (CoalHMM \cite{coalHMM1,coalHMM2,coalHMM3} and diCal \cite{diCal,Steinrucken2013}) also tackle the problem of inferring genealogical information in samples of more than two haplotypes. These methods are all derived from the coalescent with recombination, a stochastic process that encapsulates the history of a collection of DNA sequences as an ancestral recombination graph (ARG) \cite{hudson1983,seqCoal}. The hidden state associated with each genetic locus is a tree with time-weighted edges, and neighboring trees in the sequence are highly correlated with each other. Sequential changes in tree structure reflect the process of genetic recombination that slowly breaks up ancestral haplotypes over time.

The methods mentioned above all infer approximate ARGs for the purpose of demographic inference, either detecting historical changes in effective population size or estimating times of divergence and admixture between different populations or species. PSMC and CoalHMM have been used to infer ancestral population sizes in a variety of non-model organisms for which only a single genome is available \cite{groenen2012,miller2012,mailund2012,zhao2012,orlando2013,wan2013}, as well as for the Neanderthal and Denisovan archaic hominid genomes \cite{meyer2012}.  Despite this progress, the demographic inference problem is far from solved, even for extremely well-studied species like \emph{Homo sapiens} and \emph{Drosophila melanogaster} \cite{pritchard2011,gronau2011,PSMC,thornton2006,haddrill2005}. Estimates of the population divergence time between European and African humans range from 50 to 120 thousand years ago (kya), while estimates of the speciation time between polar bears and brown bears range from 50 kya to 4 million years ago \cite{hailer2012,miller2012,cahill2013}.  One reason that different demographic methods often infer conflicting histories is that they make different trade-offs between the mathematical precision of the model and scalability to larger input datasets. This is even true within the class of coalescent HMMs, which are much more similar to each other than to methods that infer demography from summary statistics \cite{dadi,palamara2012,IBS} or Markov chain Monte Carlo \cite{gronau2011}.

Exact inference of the posterior distribution of ARGs given data is a very challenging problem, the major reason being that the space of hidden states is infinite, parameterized by continuous coalescence times.  In practice, when a coalescent HMM is implemented, time needs to be discretized and confined to a finite range of values. It is a difficult problem to choose an optimal time discretization that balances the information content of a dataset, the complexity of the analysis, and the desire to infer particular periods of history at high resolution. Recent demographic history is often of particular interest, but large sample sizes are needed to distinguish between the population sizes at time points that are very close together or very close to the present.

In a coalescent HMM under a given demographic model, optimal demographic parameters can be inferred using an expectation-maximization (EM) algorithm. The speed of this EM algorithm is a function of at least three variables: the length $L$ of the genomic region being analyzed, the number $n$ of sampled haplotypes, and the number $d$ of states for discretized time.  In most cases, the complexity is linear in $L$, but the complexity in $n$ can be enormous because the number of distinct $n$-leaved tree topologies grows super-exponentially with $n$.  PSMC and CoalHMM avoid this problem by restricting $n$ to be very small, analyzing no more than four haplotypes at a time. diCal admits larger values of $n$ by using a \emph{trunk genealogy} approximation (see \cite{Paul2011,diCal,Steinrucken2013} for details) which is derived from the diffusion process dual to the coalescent process, sacrificing information about the exact structure of local genealogies in order to analyze large samples which are informative about the recent past.

To date, all published coalescent HMMs have had  quadratic complexity in $d$.  This presents a significant limitation given that small values of $d$ lead to biased parameter estimates \cite{coalHMM3} and limit the power of the method to resolve complex demographic histories. PSMC is typically run with a discretization of size $d=64$, but diCal and CoalHMM analyses of larger datasets are restricted to coarser discretizations by the cost of increasing the sample size. In this paper, we exploit the natural symmetries of the coalescent process to derive an alternate EM algorithm with linear complexity in $d$. The speedup requires no approximations to the usual forward-backward probabilities; we perform an exact computation of the likelihood in $O(d)$ time rather than $O(d^2)$ time using an augmented HMM.  We implement the algorithms presented in this paper to speed up our published method diCal, which is equivalent to PSMC when the sample size is two, yielding results of the same quality as earlier work in a fraction of the runtime. 
We have included the speedup in the most
recent version of our program diCal; source code can be
downloaded at
\url{http://sourceforge.net/projects/dical/}.

\section{Linear-Time Computation of the Forward and Backward Probabilities}
\label{sec:compute_fb}


We consider a coalescent HMM $\model$ with hidden states $\state_1, \ldots, \state_L$ and observations $\obs=\subObs{1}, \ldots, \subObs{L}$.
For PSMC, $\state_\ell$ is the discretized time interval in which two homologous chromosomes coalesce at locus $\ell$, while $\subObs\ell$ is an indicator for heterozygosity.  
The method diCal is based on the conditional sampling distribution
(CSD) which describes the probability of observing a newly sampled
haplotype $\obs$ given a collection $\cH$ of $n$
already observed haplotypes.  In diCal, the hidden state at locus
$\ell$ is $\state_\ell = (H_\ell,T_\ell)$, where $H_\ell\in\cH$
denotes the haplotype in the ``trunk genealogy'' (see \cite{Paul2011})
with which $\obs$ coalesces at locus $\ell$ and $T_\ell\in
\{1,\ldots,d\}$ denotes the discretized time interval of coalescence;
the observation $\subObs\ell \in \mathcal{A}$ is the allele of haplotype $\obs$ at locus $\ell$.
For $n=|\cH|=1$, diCal is equivalent to PSMC.  In what follows, we present our algorithm in the context of diCal, but we note that the same underlying idea can be applied to other coalescent HMMs.

\subsection{A linear-time forward algorithm \label{sec:compute_f}}
We use $f(\subObs{1:\ell},(h,j))$ to denote the joint forward probability of
observing the partial emitted sequence
$\subObs{1:\ell}:=\subObs1,\ldots,\subObs\ell$ and the hidden state $S_\ell=(h,j)$ at locus $\ell$.  
The probability of transitioning
from state $(h',k)$ at locus $\ell$ to state $(h,j)$ at locus $\ell+1$
is denoted by $\phi(h,j\mid h',k)$, 
the stationary probability of state $(h,i)$ is denoted $\zeta(h,i)$,
and the emission probability of
the observed allele $\subObs\ell=a$ given coalescence at $T_\ell=j$
onto haplotype $h$ with allele $\sub{h}{\ell}=b$ at locus $\ell$ is denoted by $\xi(a\mid b,j)$.
When $\ell$ is obvious from the context, we sometimes use $\xi(a \mid s) := \xi(a \mid \sub{h}{\ell},j)$ for $s = (h,j)$.
Explicit expressions for $\zeta(h,i)$, $\phi(h,j \mid h',k)$, and $\xi(a\mid b,j)$
in the context of our program diCal are given in \cite{diCal}.

The forward probabilities are computed using the recursion
\begin{equation}\label{standard_forward}
f(\subObs{1:\ell+1},(h,j)) =\xi(\subObs{\ell+1}|\sub{h}{\ell+1},j)\cdot \sum_{k=1}^d\sum_{h'\in\cH} f(\subObs{1:\ell},(h',k))\cdot \phi(h,j|h',k),
\end{equation} 
which contains $nd$ terms. Since there are also $nd$ possibilities for
$\state_{\ell+1} = (h,j)$, it should naively take $O(n^2d^2 L)$ time to compute the entire forward dynamic programming (DP) table $\{f(\subObs{1:\ell},\state_{\ell})\}_{\ell=1}^L$. The key to achieving a speed-up is to factor \eqref{standard_forward} in a way that reflects the structure of the coalescent, exploiting the fact that many transitions between different hidden states have identical probabilities.

\begin{figure}[t]
	\center
	\subfigure[]{
	\includegraphics[width=.54\textwidth]{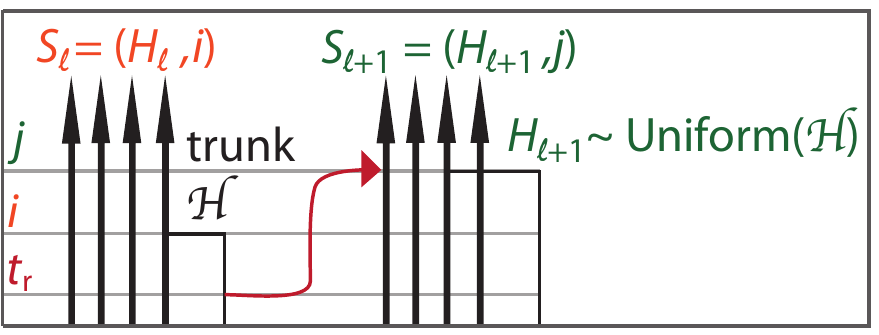} \label{fig:recomb_transition_a}}\hspace{5mm}
	\subfigure[]{
	\includegraphics[width=.36\textwidth]{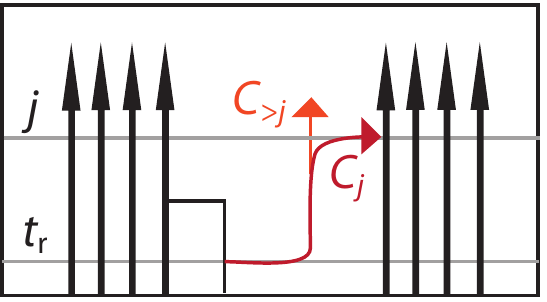} \label{fig:recomb_transition_b}}
	\caption{(a). Here, we illustrate a transition from hidden state $S_\ell=(h_n,i)$ to hidden state $S_{\ell+1}=(h_k,j)$ that proceeds via recombination at time $t_r$. The probability of this transition does not depend on the identity of the haplotype $h_k$.(b). As a recombined lineage floats through time interval $j$, it can either coalesce with the trunk (event $C_j$) or keep floating (event $C_{>j}$) and eventually coalesce with the trunk in a more ancient time interval.}
	\label{fig:recomb_transition}
\end{figure}

After a sampled lineage recombines at time $t_r$ between loci $\ell$ and $\ell+1$, it will ``float" backward in time from the recombination breakpoint until eventually coalescing with a trunk lineage chosen uniformly at random (Figure~\ref{fig:recomb_transition_a}). This implies that $\phi(h,j|h',k)=\phi(h,j|h'',k)$ whenever $h'\neq h$ and $h''\neq h$, and exploiting this symmetry allows the forward table to be computed in $O(nd^2L)$ time. This speed-up was already implemented in the algorithm described in Paul \emph{et al.}~\cite{Paul2011}.

Another symmetry of the transition matrix, not exploited previously, can be
found by decomposing the transition from locus $\ell$ to locus $\ell+1$
as a sequence of component events.
In particular, let  
$R_i$ be the event that
a recombination occurs during time
interval $i$,
and let $\overline{R}$ be the
event that no recombination occurs between $\ell$ and $\ell+1$.
Then we have that
\begin{align}
  \phi((h,j) \mid (h',k)) 
&= 
\frac1n \sum_{i=1}^{\min(j, k)} \left( \mathbb{P}(R_i, T_{\ell+1} = j \mid T_\ell = k) \right. \notag \\
&+ \left.\bbOne_{\{(h,j)=(h',k')\}} \mathbb{P}(\overline{R} \mid T_\ell = k)\right), \label{transition:prob}
\end{align}
where $\bbOne_{E}=1$ if the event $E$ is true or $0$ otherwise.
The factor $1/n$  corresponds to the probability that the sampled lineage coalesces with haplotype $h\in\cH$ in the trunk genealogy.

If a recombination occurs in time interval $i$, the sampled lineage will start to ``float" freely back in time until it either coalesces in $i$ or floats into the next time interval $i+1$ (Figure~\ref{fig:recomb_transition_b}).
Specifically, we let 
$C_{>i}$ denote the event where the sampled lineage 
recombines at or before $i$ and
floats
into $i+1$,
and
$C_i$ denote the event where the recombined lineage
coalesces back in interval $i$.
Noting that $\mathbb{P}(R_i, C_i \mid T_\ell =i')$ and
$\mathbb{P}(R_i, C_{>i} \mid T_\ell=i')$
are independent of $i'$ whenever $i' > i$,
and that coalescence happens as a Markov process backwards in time, we obtain
\begin{align}\label{prob:recom}
\bbP(R_i, T_{\ell+1} = j & \mid T_\ell = k) = \bbOne_{i=j=k}\cdot   \mathbb{P}(R_i, C_i \mid T_\ell = i)  \notag \\
&+ \bbOne_{i=j<k} \cdot \mathbb{P}(R_i, C_i \mid T_\ell > i) \notag \\
&+ \bbOne_{i=k<j}\cdot  \mathbb{P}(R_i, C_{>i} \mid T_\ell = i)\cdot \prod_{k=i}^{j-1} \mathbb{P}(C_{>k+1} \mid C_{>k}) \notag \\
&+\bbOne_{i<\min(j,k)} \cdot  \mathbb{P}(R_i, C_{>i} \mid T_\ell > i)\cdot \prod_{k=i}^{j-1} \mathbb{P}(C_{>k+1} \mid C_{>k}).
\end{align}
Explicit formulas (specific to the method diCal) for the above terms are provided in the appendix.

By combining \eqref{transition:prob} with \eqref{prob:recom}
and then collecting terms in \eqref{standard_forward},
we can remove the sum over $T_\ell = k$ when computing
$f(\subObs{1:\ell+1}, \state_{\ell+1})$.
In particular, 
we
define additional forward probabilities
\begin{align}
  \feq{k}{\ell} &:= \mathbb{P}(\subObs{1:\ell}, T_\ell = k) =\sum_{h'\in\cH} f(\subObs{1:\ell},S_{\ell}=(h',k)), \label{feq}\\
  \fsum{k}{\ell} &:= \mathbb{P}(\subObs{1:\ell}, T_\ell > k) =\sum_{k'=k+1}^d\sum_{h'\in\cH} f(\subObs{1:\ell}, S_{\ell}=(h',k')), \label{fsum}\\
\fgreaterthan{j}{\ell} &:= \sum_{i=1}^j \mathbb{P}(\subObs{1:\ell}, R_i, \float{i}, \ldots, \float{j}) \label{extra_forward}
\\
&
\phantom{:}= \sum_{i=1}^j
  \Bigg\{\bigg[ \prod_{i'=i+1}^{j} \mathbb{P}(\float{i'} \mid \float{i'-1}) \bigg] 
\notag\\  
\times 
\bigg[\feq{i}{\ell} &\mathbb{P}(R_i, \float{i} \mid T_\ell = i)
+ \fsum{i}{\ell} \mathbb{P}(R_i, \float{i}\mid T_{\ell}>i)\bigg] \Bigg\} . \notag
\end{align}
Then, \eqref{standard_forward} can be written as 
\begin{eqnarray}\label{final_forward}
f(\subObs{1:\ell+1},(h,j)) &=& \xi(\subObs{\ell+1}\mid \sub{h}{\ell+1},j)\cdot\bigg[ 
\frac{1}{n} \fgreaterthan{j-1}{\ell}\mathbb{P}(\coal{j}|\float{j-1})\notag \\
&& + \frac{1}{n}\fsum{j}{\ell} \mathbb{P}(R_j,C_j \mid T_\ell > j) \notag \\
&& +\frac{1}{n}\feq{j}{\ell} \mathbb{P}(R_j,\coal{j}\mid T_{\ell}=j) \notag\\
&& + f(\subObs{1:\ell},(h,j)) \mathbb{P}(\overline{R}\mid T_{\ell}=j)
\bigg].
\end{eqnarray}
This can be seen by noting that the first three terms in the sum correspond
to the terms for $i < j$, $i=j < k$, and $i=j=k$, respectively when
putting together
\eqref{standard_forward}
and \eqref{transition:prob}.
Alternatively, \eqref{final_forward} follows from directly considering the
probabilistic interpretation of the terms $f(\subObs{1:\ell},*)$ as
given by \eqref{feq}, \eqref{fsum}, and \eqref{extra_forward}.

The required values of $\fgreaterthan{i}{\ell}$ and $\fsum{i}{\ell}$ can be computed recursively using 
\allowdisplaybreaks
\begin{eqnarray}
\fsum{i}{\ell}&=&\fsum{i+1}{\ell}+\feq{i+1}{\ell}, \label{final_fsum} \\
\fgreaterthan{i}{\ell} &=& \fgreaterthan{i-1}{\ell}\mathbb{P}(\float{i}|\float{i-1})  \nonumber\\
&&+\feq{i}{\ell} \mathbb{P}(R_i,\float{i}\mid T_\ell = i) \nonumber\\
&&+
\fsum{i}{\ell}\mathbb{P}(R_i,\float{i}\mid T_\ell > i),\label{final_fgreaterthan}
\end{eqnarray}
with the base cases
\begin{eqnarray*}
\fsum{d}{\ell} &=& 0, \\
\fgreaterthan{1}{\ell} &=& \fsum{1}{\ell}\mathbb{P}(R_1,\float{1}\mid T_\ell > 1) \notag \\
&&+\feq{1}{\ell}\mathbb{P}(R_1,\float{1}\mid T_\ell = 1).
\end{eqnarray*}
Hence, using the recursions \eqref{final_forward}, ~\eqref{final_fsum}, and~\eqref{final_fgreaterthan},  it is possible to compute 
the entire forward DP table  $\{f(\subObs{1:\ell},\state_{\ell})\}_{\ell=1}^L$ exactly in $O(ndL)$ time.

\subsection{A linear-time backward algorithm}
The backward DP table $\{b(\subObs{\ell+1:L}\mid \state_\ell)\}$ can be also computed in $O(ndL)$ time.  
Given the linear-time forward algorithm discussed in the previous section, the easiest way to compute the backward DP table is as follows:  Let $\revObs=\subRevObs{1},\subRevObs{2},\ldots,\subRevObs{L}=\subObs{L},\subObs{L-1},\ldots,\subObs{1}$ denote the reversed $x$ and let $S^{(r)}_\ell$ denote the hidden states for the HMM generating $\revObs$. 
Then, since the coalescent is reversible along the sequence, 
\begin{eqnarray*}
b(\subRevObs{\ell+1:L}\mid s) 
=\frac{\mathbb{P}(\subRevObs{\ell+1:L}, S_\ell=s)}{\zeta(s)}=\frac{\bbP(\subRevObs{\ell:L},S_\ell=s)}{\xi(\subRevObs\ell\mid s)\zeta(s)}=\frac{f(\subRevObs{1:L-\ell+1}, S^{(r)}_{L-\ell+1}=s)}{\xi(\subRevObs\ell\mid s)\zeta(s)}.
\end{eqnarray*}


\section{Linear-Time EM via an Augmented HMM}

The primary application of PSMC and diCal is
parameter estimation, specifically the estimation of demographic
parameters such as changing population sizes.
This is done through a maximum likelihood framework with the
expectation maximization (EM) algorithm.
In this section, we describe how to speed up the EM algorithm to work
in linear time.

\subsection{The standard EM algorithm with $O(d^2)$ time complexity}
Let $\Theta$ denote the parameters we wish to estimate,
and $\hat{\Theta}$ denote the maximum likelihood estimate:
\begin{align*}
  \hat{\Theta} = \arg \max_{\Theta'} \mathcal{L}(\Theta') = \arg
  \max_{\Theta'} \mathbb{P}_{\Theta'}(\subObs{1:L}).
\end{align*}
To find $\hat{\Theta}$, we pick some initial value $\Theta^{(0)}$,
and then iteratively solve for $\Theta^{(t)}$ according to
\begin{align*}
  \Theta^{(t)} = \arg \max_{\Theta'} \mathbb{E}_{\state_{1:L}; \Theta^{(t-1)}} [ \log
  \mathbb{P}_{\Theta'}(\subObs{1:L}, \state_{1:L} )  \mid
    \subObs{1:L}],
\end{align*}
where $S_{1:L} := S_1,\ldots,S_L$.
The sequence $\Theta^{(0)},\Theta^{(1)},\ldots$ is then guaranteed to
converge
to a local maximum of the surface $\mathcal{L}(\Theta)$.

Since $(\subObs{1:L}, \state_{1:L})$ forms an HMM, the joint likelihood $\mathbb{P}(\subObs{1:L}, \state_{1:L})$ can be written as
\begin{align*}
  \mathbb{P}_{\Theta'}(\subObs{1:L}, \state_{1:L}) &= \zeta_{\Theta'}(\state_1)
  \left[ \prod_{\ell=1}^L \xi_{\Theta'}(\subObs\ell \mid \state_\ell) \right]
  \left[\prod_{\ell=2}^L \phi_{\Theta'}(\state_\ell \mid \state_{\ell-1}) \right].
\end{align*}
Letting $\bbE[\# \ell: E \mid \subObs{1:L}]$ denote the posterior
expected number of loci where event $E$ occurs,
and 
$\pi(\obs) := \mathbb{P}(\obs) =  \sum_{s} f(\subObs{1:L}, s)$
denote the total probability of observing $\obs$,
we then have
\begin{align}
  & \mathbb{E}_{\state_{1:L}; \Theta} \big[ \log
  \mathbb{P}_{\Theta'}(\subObs{1:L}, \state_{1:L} )  \big|
    \subObs{1:L} \big] \notag\\
    = &
    \sum_{s} \left(\log \zeta_{\Theta'}(s)\right)
    \mathbb{P}_{\Theta}(S_1 = s | \subObs{1:L}) \notag \\
  &\qquad + \sum_{(h,i)} \sum_{a,b \in \mathcal{A}} \left(\log \xi_{\Theta'} (a | b,i) \right)
  \mathbb{E}_\Theta\big[\# \ell: \{\state_\ell = (h,i), \sub{h}{\ell} = b, \subObs\ell = a\} \big| \subObs{1:L} \big]
 \notag\\
  & \qquad
    +  \sum_{s,s'} \left(\log \phi_{\Theta'}(s' \mid s)\right)
    \mathbb{E}_\Theta\big[\# \ell : \{\state_{\ell-1} = s, \state_\ell = s'\} \big| \subObs{1:L}  \big]
 \notag\\
    = &
 \frac1{\pi_\Theta(x)} \Bigg[    \sum_{s} \left(\log \zeta_{\Theta'}(s)\right)
    f_\Theta(\subObs1, s) b_\Theta(\subObs{2:L} | s) \notag \\
&\qquad  + \sum_{(h,i)} \sum_{a,b \in \mathcal{A}} \left(\log \xi_{\Theta'} (a| b,i) \right)\!\!
  \sum_{\substack{\ell : \subObs\ell = a \\ \sub{h}{\ell} = b}} \!\! f_\Theta(\subObs{1:\ell}, (h,i))
    b_\Theta(\subObs{\ell+1:L} | h,i)   \notag\\
  & 
    +  \sum_{s,s'} \left(\log \phi_{\Theta'}(s' \mid s)\right)
    \left(\sum_{\ell=1}^{L-1} f_{\Theta}(\subObs{1:\ell}, s) \phi_\Theta(s'
      \mid s) \xi_\Theta(\subObs{\ell+1} \mid s') b_\Theta(\subObs{\ell+2:L} \mid s') \right)\Bigg]. \label{em:objective}
\end{align}
Note that we have to compute the term $\sum_{\ell} f_{\Theta}(\subObs{1:\ell}, s) \phi_\Theta(s'\mid s) \xi_\Theta(\subObs{\ell+1} \mid s')$ $b_\Theta(\subObs{\ell+2:L} \mid s')$
for every pair of states $s,s'$, which makes computing the EM
objective function
quadratic in the number $d$ of discretization time intervals, despite the fact that we
computed
the forward and backward tables in linear time.

\subsection{A linear-time EM algorithm}
By augmenting our HMM to condition on whether
recombination occurred between loci $\ell$ and $\ell+1$,
the EM algorithm can be sped up to be linear in $d$. We now describe this augmented HMM.
Let $\model$ denote our original HMM, with states $\state_{1:L}$ and
observations $\subObs{1:L}$.
Between loci $\ell$ and $\ell+1$, define
\begin{equation}
  \recom_{l,l+1} = \begin{cases}
    \overline{R},   &\text{if no recombination,}\\
    R_i,   &\text{if recombination occurred at time } i.
    \end{cases}\notag
\end{equation}
Now let $\state^*_1 = \state_1$, and $\state^*_\ell =
(\recom_{\ell-1,\ell}, \state_\ell)$ for $\ell > 1$.
We let $\model^*$ be the HMM with hidden variables $\state^*_{1:L} =
\state^*_1,\ldots,\state^*_L$,
observations $\subObs{1:L}$,
transition probabilities $\mathbb{P}(\state^*_\ell \mid \state^*_{\ell-1}) =
\mathbb{P}(\state^*_\ell \mid \state_{\ell-1})$,
and emission probabilities $\mathbb{P}(\subObs{\ell} \mid \state^*_\ell) =
\mathbb{P}(\subObs{\ell} \mid \state_\ell)$.
Note that the probability of observing the data is the same under $\model$
and $\model^*$, i.e.,
\begin{align*}
    \mathcal{L}(\Theta) &= \mathbb{P}_{\Theta}(\subObs{1:L}\mid \model) =
    \mathbb{P}_{\Theta}(\subObs{1:L}\mid \model^*),
\end{align*}
and so we may find a local maximum of $\mathcal{L}(\Theta)$ by
applying the EM algorithm to the augmented HMM $\model^*$, instead of
to
the original HMM $\model$.

\begin{figure}[t]
	\center
	\includegraphics[width=.45\textwidth]{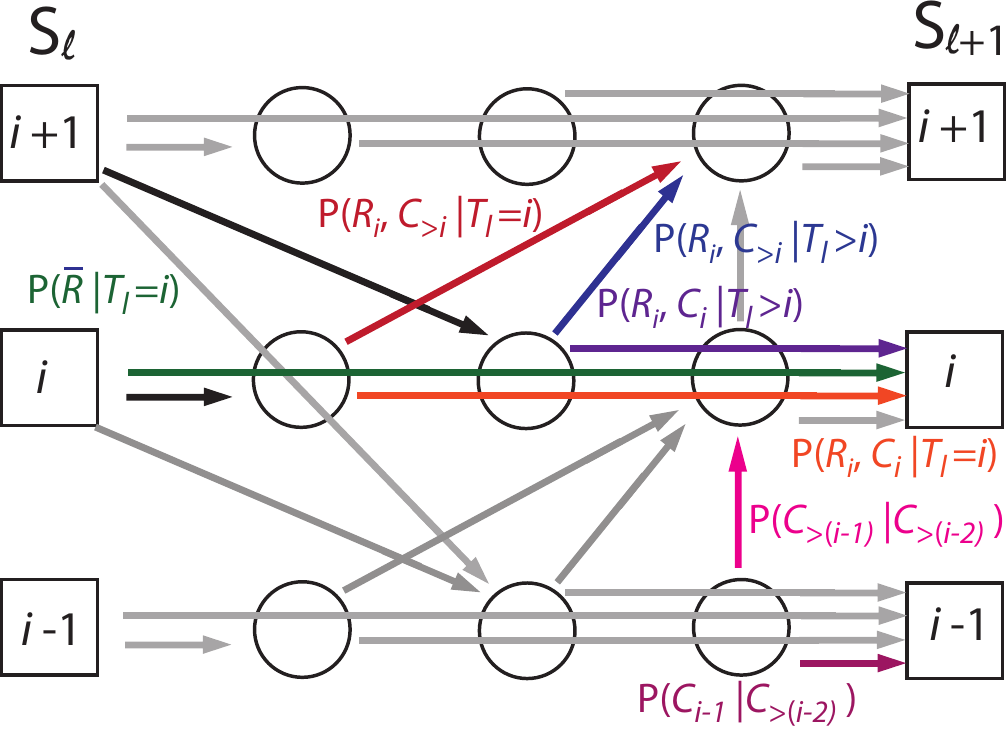}
	\caption{ This diagram illustrates the flow of transition probabilities through the augmented HMM. Lineages may transition between different coalescence times at loci $\ell$ and $\ell+1$ by recombining and passing through the floating states represented by circles. Each interval contains three distinct floating states to capture the the dependence of recombination and coalescence probabilities on whether any of these events occur during the same time interval.}
	\label{fig:augmented_HMM}
\end{figure}

To compute the EM objective function for $\model^*$,
we start by noting that the joint likelihood is
\begin{align}
\mathbb{P}(\subObs{1:L}, \state^*_{1:L})   & =
  \zeta(\state_1) \bigg[\prod_{\ell=1}^L \xi(\subObs\ell \mid
    \state_\ell) \bigg]
\bigg[ \prod_{\substack{\ell: \recom_{\ell,\ell+1} = \overline{R}}}
\mathbb{P}(\overline{R}\mid T_{\ell} ) \bigg] \label{joint:likelihood:augmented} \\
& \qquad \qquad \times
 \bigg[ \prod_{i=1}^d 
\prod_{\substack{\ell:  \recom_{\ell,\ell+1} = R_i}}
 \bbP(R_i, T_{\ell+1} \mid
T_\ell) \bigg] \left(\frac1{n}\right)^{\# \ell: \recom_{\ell,\ell+1} \neq \overline{R}},  \notag
\end{align}
where we decomposed the joint likelihood into the initial
probability, the emission probabilities,
the transitions without recombination, and the transitions with
recombination. 
We note that the initial probability can be decomposed as
\begin{align}
  \zeta(S_1 = (h,j)) &= \frac1n \Bigg[ \prod_{i=1}^{j-1}
    \mathbb{P}(C_{>i} \mid C_{>i-1}) \Bigg]
  \mathbb{P}(C_j \mid C_{>j-1}), \label{marginal:decomposition}
\end{align}
and from \eqref{prob:recom}, we
decompose the product of transition recombination probabilities as
\allowdisplaybreaks
\begin{align}
\prod_{i=1}^d \prod_{\substack{\ell:  \recom_{\ell,\ell+1} = R_i}}
\bbP(R_i, T_{\ell+1} \mid
T_\ell)
&= 
 \prod_{i=1}^d\Bigg\{
\bigg[\prod_{\substack{\ell:  \recom_{\ell,\ell+1} = R_i \\ T_\ell =
    T_{\ell+1} = i}} \mathbb{P}(R_i, C_i \mid T_\ell = i) \bigg]
\notag\\ 
& \hspace{-40mm} \qquad \times  \bigg[\prod_{\substack{\ell:  \recom_{\ell,\ell+1} = R_i \\ T_\ell >
    T_{\ell+1} = i}} \mathbb{P}(R_i, C_i \mid T_\ell > i) \bigg] 
\bigg[\prod_{\substack{\ell:  \recom_{\ell,\ell+1} = R_i \notag \\
    T_{\ell+1} > T_\ell = i}} \mathbb{P}(R_i, C_{>i} \mid T_\ell = i) \bigg] \\
& \hspace{-40mm} \qquad \times
\bigg[\prod_{\substack{\ell:  \recom_{\ell,\ell+1} = R_i \\ 
     T_\ell, T_{\ell+1} > i}} \mathbb{P}(R_i, C_{>i} \mid T_\ell > i) \bigg] 
     \bigg[ \prod_{\substack{\ell: T_{\ell+1} > i \\ \recom_{\ell,\ell+1}
  \in R_{<i}}} \mathbb{P}(C_{>i} \mid C_{>i-1}) \bigg] 
\notag\\ 
& \hspace{-40mm}  \qquad \times
\bigg[ \prod_{\substack{\ell: T_{\ell+1} = i \\ \recom_{\ell,\ell+1}
  \in R_{<i}}} \mathbb{P}(C_{i} \mid C_{>i-1}) \bigg] \Bigg\}, \label{recom:transition:expanded}
\end{align}
where $R_{<i} := \cup_{j <i} R_j$.
Figure~\ref{fig:augmented_HMM} shows a graphical representation for the transitions
of $\model^*$.

By plugging \eqref{marginal:decomposition} and \eqref{recom:transition:expanded} into \eqref{joint:likelihood:augmented},
then taking the posterior expected logarithm of
\eqref{joint:likelihood:augmented}, we obtain the EM objective
function for $\model^*$:
\begin{align}
  \mathbb{E}_{\state^*_{1:L}; \Theta} \Big[ \log
  \mathbb{P}_{\Theta'}(\subObs{1:L}, \state^*_{1:L} )  \mid
    \subObs{1:L}\Big]
    &= - L \log n + \sum_{i=1}^d q_i(\Theta, \Theta'), \label{em:objective:augmented}
\end{align}
where        
\allowdisplaybreaks
\begin{align}
q_i&(\Theta,\Theta')
     := 
  \sum_{a,b \in \mathcal{A}} 
  \Bigg[\frac{\log \xi_{\Theta'} (a \mid b,i)}{\pi_\Theta(x)} \sum_{\substack{\ell : \subObs\ell = a }} \sum_{ h: \sub{h}{\ell} = b} f_\Theta(\subObs{1:\ell}, (h,i))
    b_\Theta(\subObs{\ell+1:L} \mid (h,i)) \Bigg] 
\notag\\
  & 
  +   \left(\log \mathbb{P}_{\Theta'} (\overline{R} \mid
    T = i)
  + \log n\right)
 \mathbb{E}_\Theta \Big[ \# \ell: \{\recom_{\ell,\ell+1} = \overline{R}, T_\ell = i \} \mid
 \subObs{1:L} \Big]
   \notag\\
  & 
  +  \left( \log \mathbb{P}_{\Theta'}(R_i, C_i \mid T
    = i) \right)
  \mathbb{E}_\Theta\Big[ \# \ell:
  \{\recom_{\ell,\ell+1} = R_i, T_\ell = T_{\ell+1} = i\} \mid \subObs{1:L}  \Big] \notag\\
  & 
  +  \left( \log \mathbb{P}_{\Theta'}(R_i, C_i \mid T >i) \right)
  \mathbb{E}_\Theta\Big[ \# \ell:
  \{ \recom_{\ell,\ell+1} = R_i, T_\ell > T_{\ell+1} = i \} \mid \subObs{1:L}  \Big] \notag\\
  & 
  +  \left( \log \mathbb{P}_{\Theta'}(R_i, C_{>i} \mid T= i) \right)
  \mathbb{E}_\Theta\Big[ \# \ell:
  \{\recom_{\ell,\ell+1} = R_i, T_{\ell+1} > T_{\ell} = i\}\mid \subObs{1:L}  \Big] \notag\\
  & 
  +  \left( \log \mathbb{P}_{\Theta'}(R_i, C_{>i} \mid T >i) \right)
  \mathbb{E}_\Theta\Big[ \# \ell:
  \{\recom_{\ell,\ell+1} = R_i, T_\ell>i , T_{\ell+1} > i\}\mid \subObs{1:L}  \Big] \notag\\
&  +  \left( \log \mathbb{P}_{\Theta'}(C_{>i} \mid C_{>i-1}) \right)
  \mathbb{E}_\Theta\Big[ \# \ell:
  \{\recom_{\ell,\ell+1} \in R_{<i}, T_{\ell+1} > i\}\mid \subObs{1:L}
  \Big] 
\notag\\
  & 
  +  \left( \log \mathbb{P}_{\Theta'}(C_{i} \mid C_{>i-1}) \right)
  \mathbb{E}_\Theta\Big[ \# \ell:
  \{\recom_{\ell,\ell+1} \in R_{<i}, T_{\ell+1} = i \} \mid \subObs{1:L} 
   \Big] \notag\\
& + \mathbb{P}_\Theta(T_1 > i \mid \subObs{1:L})  + \mathbb{P}_\Theta(T_1 = i \mid \subObs{1:L}).
\label{em:objective:augmented:sub}
\end{align}
The computation time for each of the posterior expectations
$\mathbb{E}_\Theta[\# \ell : * \mid \subObs{1:L} ]$
and $\mathbb{P}_\Theta(T_1 \mid \subObs{1:L})$
does not depend on $d$; full expressions
are listed in the appendix.
Hence, the number of operations needed to evaluate
\eqref{em:objective:augmented} is
linear in $d$.

We note another attractive property of
\eqref{em:objective:augmented}.
By decomposing the EM objective function into
a sum of
terms $q_i(\Theta,\Theta')$,
we obtain
a natural strategy for searching
through the parameter space.
In particular, one can attempt to find
the $\arg \max_{\Theta'}$ of
 \eqref{em:objective:augmented}
by optimizing the $q_i(\Theta,\Theta')$
one at a time in $i$.
In fact, for the problem of estimating changing population sizes,
$q_i(\Theta,\Theta')$ depends on $\Theta'$
almost entirely through a single parameter
(the population size $\lambda_i'$ in interval $i$),
and we pursue a strategy of
iteratively solving for $\lambda_i'$
while holding the other coordinates of $\Theta'$ fixed,
thus reducing a multivariate optimization problem into a sequence
of univariate optimization problems.

Although both the linear and quadratic EM procedures are guaranteed to converge
to local maxima of $\mathcal{L}(\Theta)$, they may have
different rates of convergence, and may converge to different
local maxima.
The search paths of the two EM algorithms may differ for two reasons:
first, the intermediate objective functions \eqref{em:objective}
and \eqref{em:objective:augmented}
are not equal,
and secondly,
as discussed above,
we use different search strategies
to find the optima of \eqref{em:objective} and
\eqref{em:objective:augmented}. We have no proven guarantee that
either search should perform better than the other, but our
observations indicate that the linear-time EM algorithm
typically converges to a value of
$\Theta$ with a equal or higher value
of $\mathcal{L}(\Theta)$ than
the quadratic-time algorithm,
in a fraction of the time
(see Figure~\ref{fig:convergence} for an example).

\section{Results}

\begin{figure}[t]
	\center
	\subfigure[]{
	\includegraphics[width=.47\textwidth]{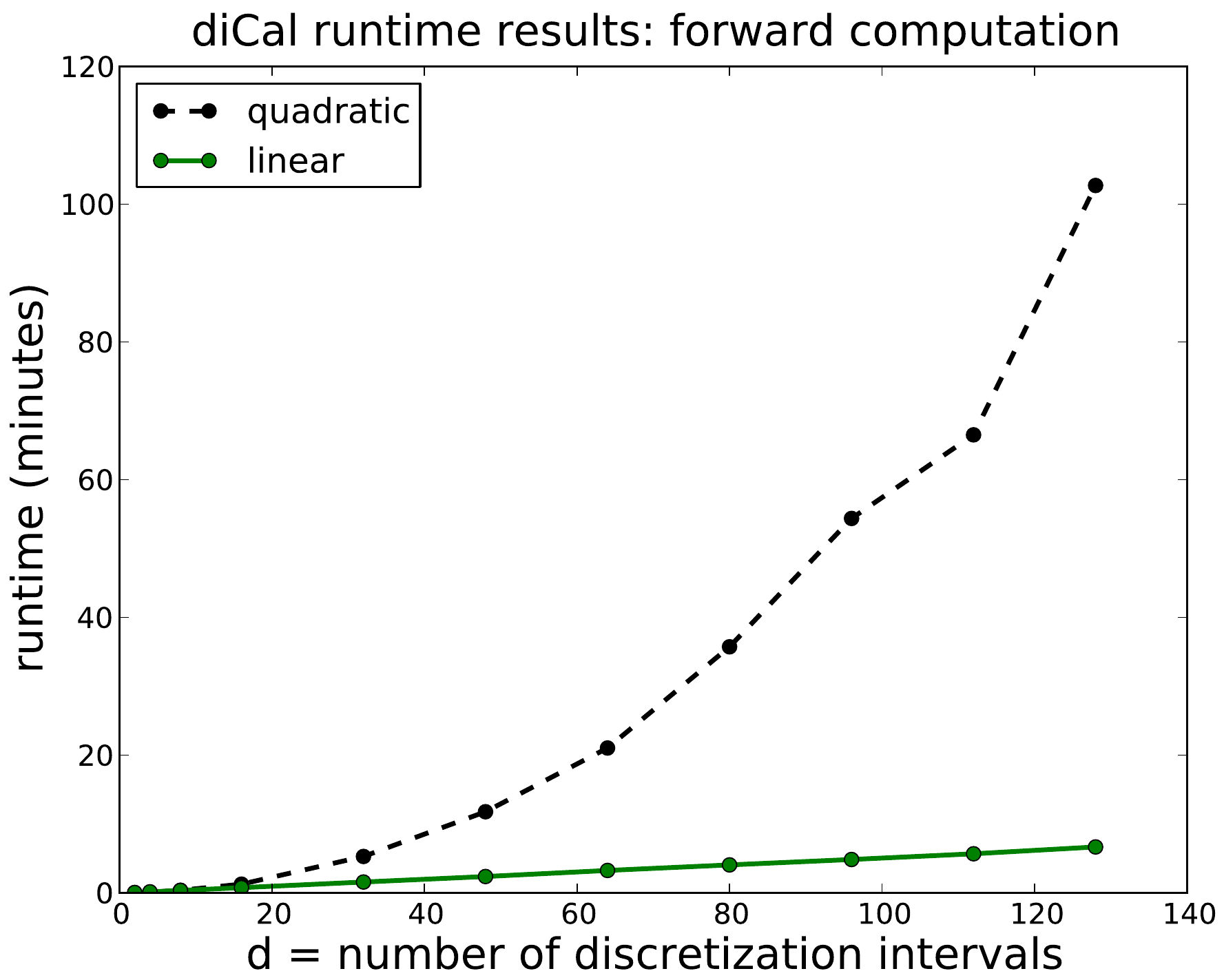}
	\label{fig:runtime_forward}}
	\subfigure[]{
	\includegraphics[width=.47\textwidth]{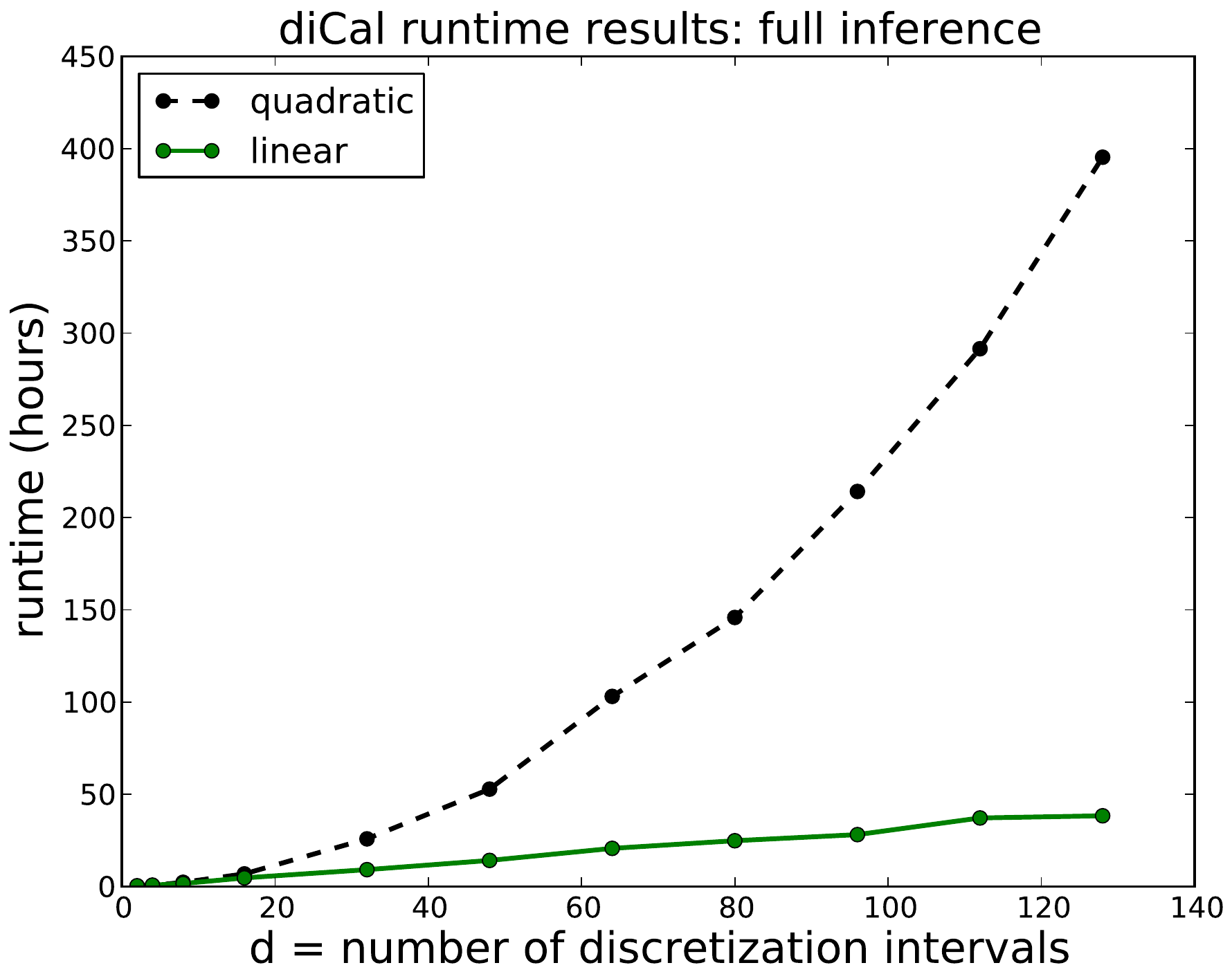}
	\label{fig:runtime_EM}}
	\caption{Runtime results on simulated data with $L=2$~Mb and $2$ haplotypes, for varying number $d$ of discretization intervals. 
\subref{fig:runtime_forward} Runtime results (in minutes) for the forward computation.
\subref{fig:runtime_EM} Runtime results (in hours) for the entire EM inference algorithm (20 iterations) extrapolated from the time for one iteration.
	}
	\label{fig:runtime}
\end{figure}

To confirm the decrease in runtime, we ran the linear-time diCal method on simulated data with $L=2$~Mb of loci and  $2$ haplotypes (in which case diCal is equivalent to PSMC), using $d=2, 4, 8, 16, 32, 48, 64, 80, 96, 112, 128$ discretization intervals.  To simulate the data, we used \verb=ms= \cite{hudson2002} with a population-scaled recombination rate $\rho=0.0005$ to generate an ARG, and then added mutations using a population-scaled mutation rate of $\theta=0.0029$ and a finite-sites mutation matrix described in Sheehan \emph{et al.}~\cite{diCal}.    Figure~\ref{fig:runtime_forward} shows the time required to compute the table of forward probabilities.  We also measured the time required for one EM iteration and then extrapolated to 20 iterations to approximate the time required to estimate an effective population size history (Figure~\ref{fig:runtime_EM}). 
In both figures, the linear runtime of our new algorithm is apparent and significantly improves our ability to increase the number of discretization intervals.


To assess the gain in accuracy of population size estimates that is
afforded by more discretization intervals, we ran both the linear- and
quadratic-time methods on simulated data with $10$ haplotypes and
$L=2$~Mb.  The conditional sampling distribution was used in a
leave-one-out composite likelihood approach \cite{diCal} in this
experiment. To run each method for roughly the same amount of time
($\approx 40$ hours), we used $d=9$ for the quadratic method and
$d=21$ for the linear method.
For both methods, we ran the EM for 20 iterations and inferred $d/3$ size change parameters.
As measured by the PSMC error function, which integrates the absolute value of the difference between the true size function and the estimated size function \cite{PSMC}, larger values of $d$ permit the inference of more accurate histories.

\begin{figure}[t]
	\center
	\subfigure[]{
	\includegraphics[width=.44\textwidth]{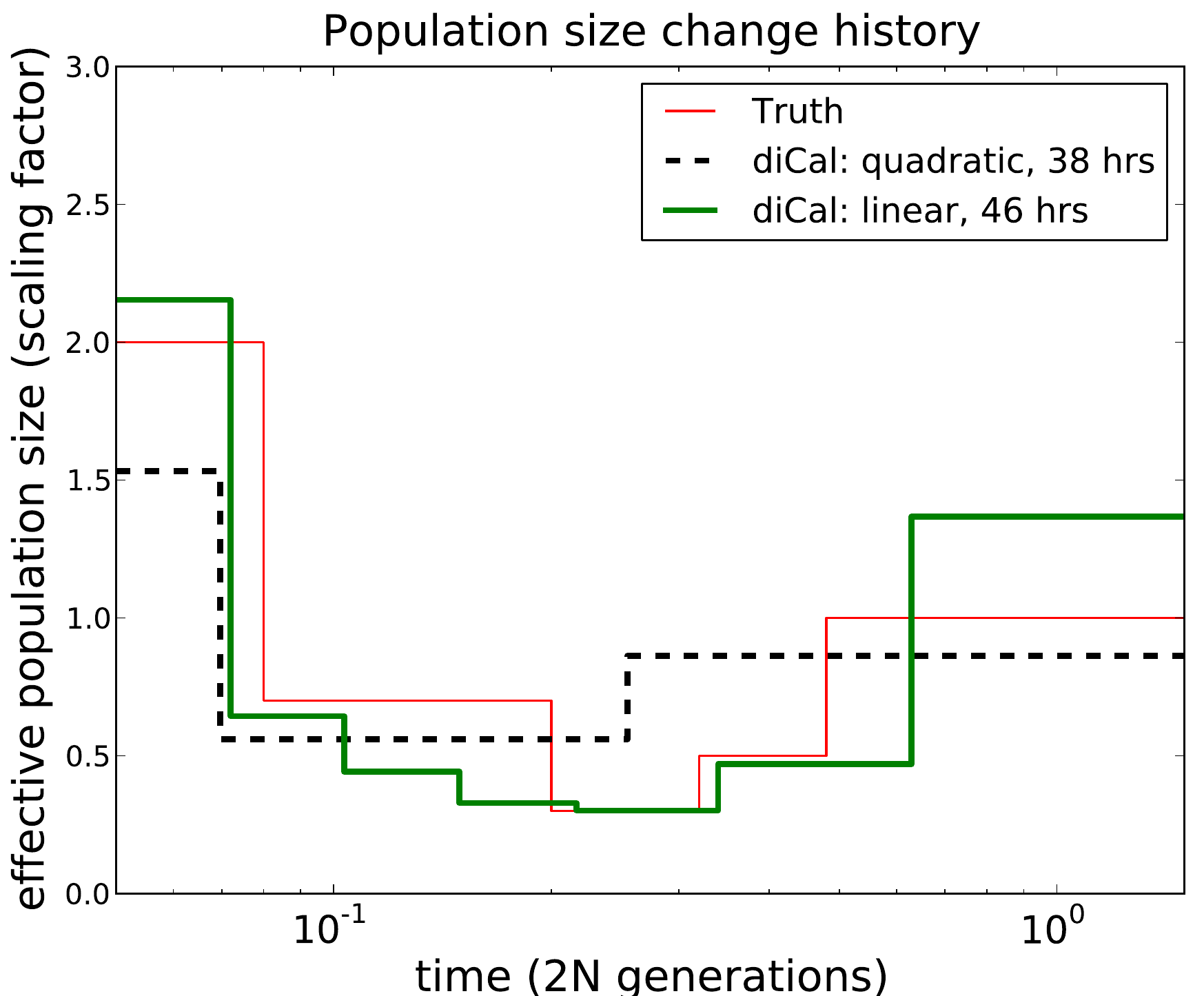}
	\label{fig:simulated_results}}
	\hspace{3mm}
	\subfigure[]{
	\includegraphics[width=.46\textwidth]{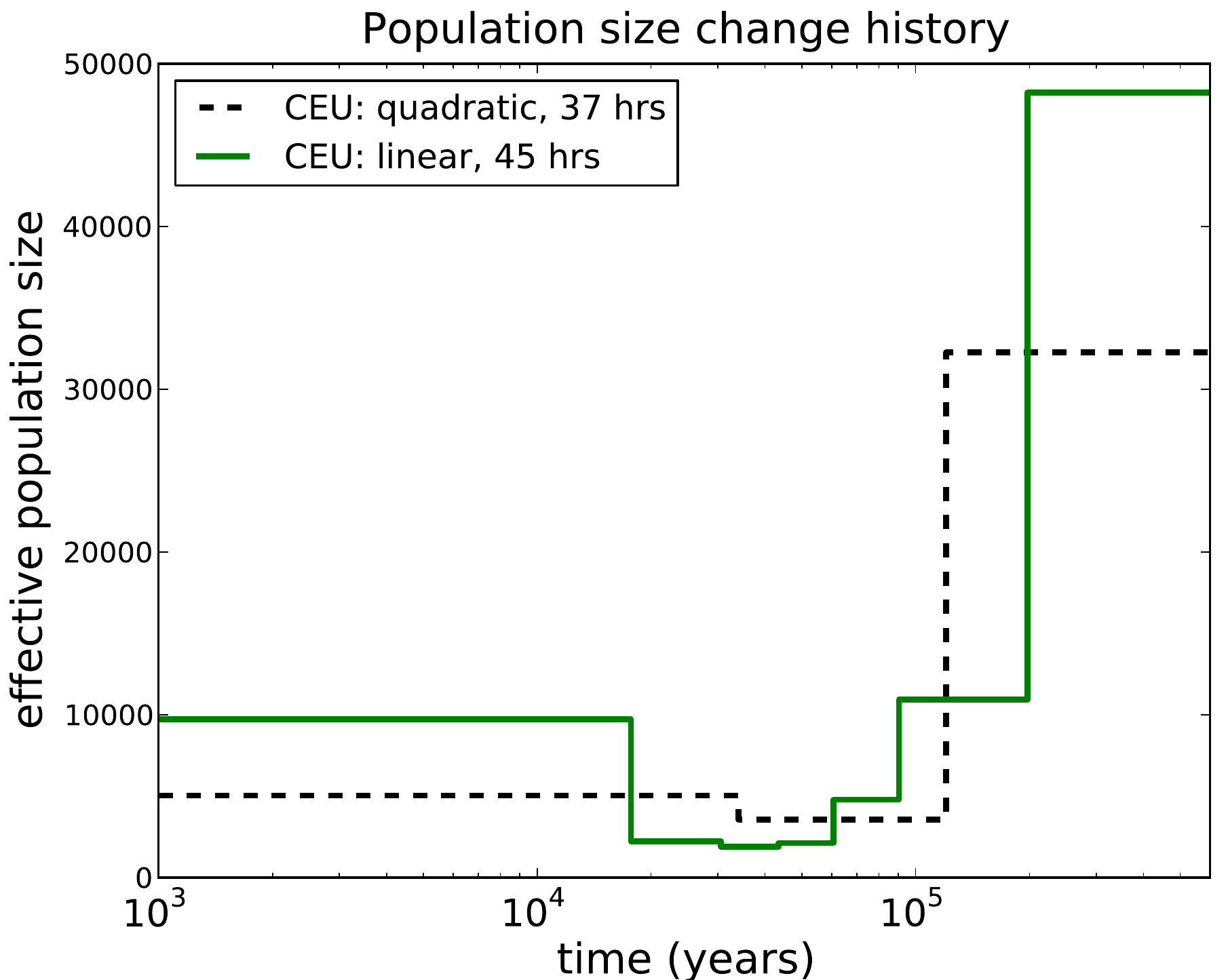}
	\label{fig:real_results}}
	\caption{Effective population size change history results.  The speedup from the linear method allows us to use a finer discretization ($d=21$) than the quadratic method ($d=9$) for about the same amount of runtime.   \subref{fig:simulated_results} Results on simulated data with $L=2~$Mb and $10$ haplotypes. Using the quadratic method with $d=9$, the error was 0.148. Using the linear method with $d=21$, the error dropped to 0.079.  \subref{fig:real_results} Results on $10$ European haplotypes over a $2$~Mb region of chromosome 1. The out-of-Africa bottleneck is very apparent with $d=21$, but is not as well characterized for $d=9$.
	}
	\label{fig:results}
\end{figure}

We also ran our method on $10$ CEU haplotypes (Utah residents of European descent) sequenced during Phase I of the the 1000 Genomes Project \cite{1000genomes} (Figure~\ref{fig:real_results}). We can see that for the quadratic method with $d=9$, we are unable to fully characterize the out-of-Africa bottleneck. In the same amount of computational time, we can run the linear method with $d=21$ and easily capture this feature. The disagreement in the ancient past between the two methods is most likely due to diCal's lack of power in the ancient past when there are not many coalescence events.  Using a leave-one-out approach  with $10$ haplotypes, the coalescence events in the ancient past tend to be few and unevenly spaced, resulting in a less confident inference.

\begin{figure}[t]
	\center
	\subfigure[]{
	\includegraphics[width=.4625\textwidth]{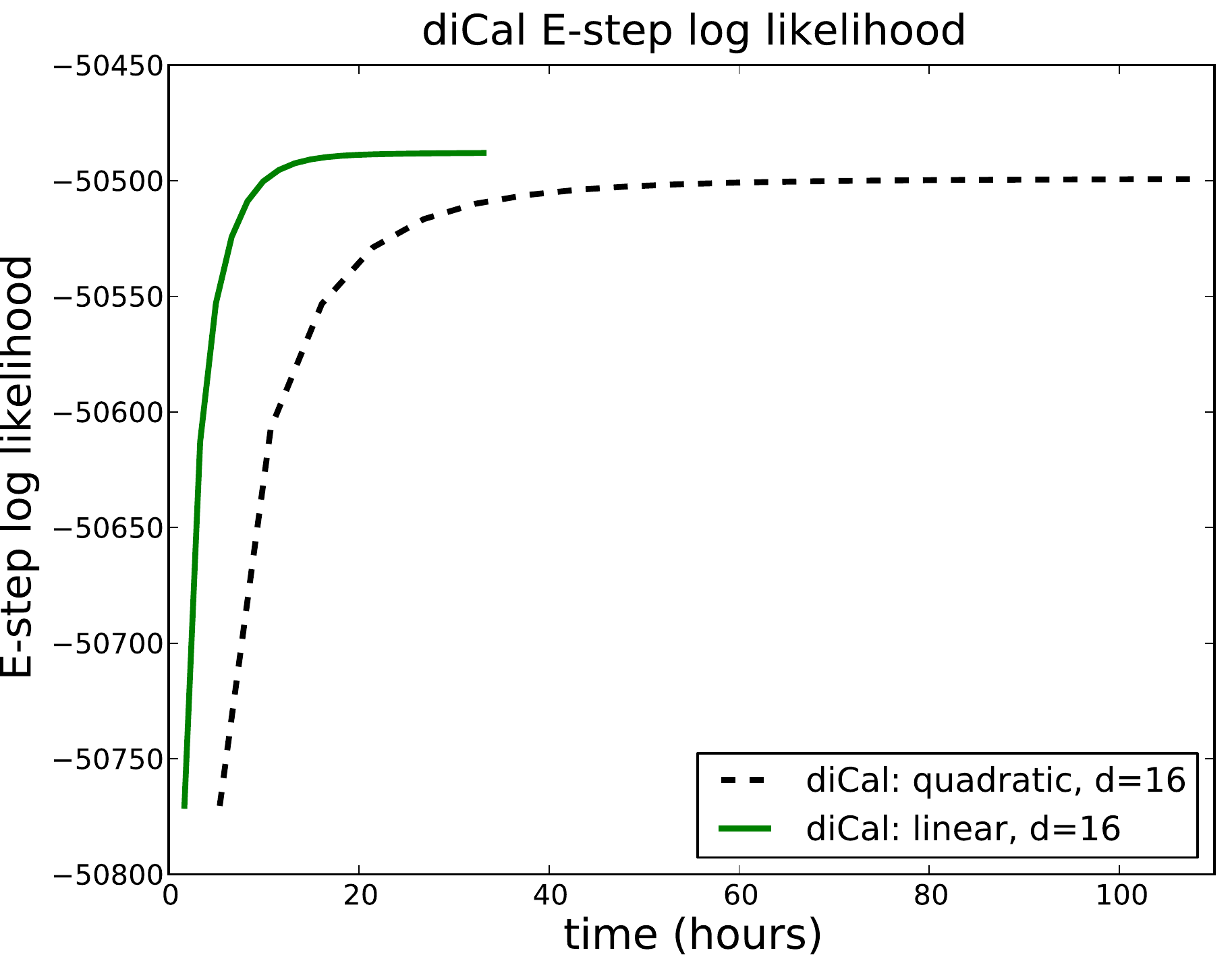}
	\label{fig:likelihood}}
	\hspace{3mm}
	\subfigure[]{
	\includegraphics[width=.4375\textwidth]{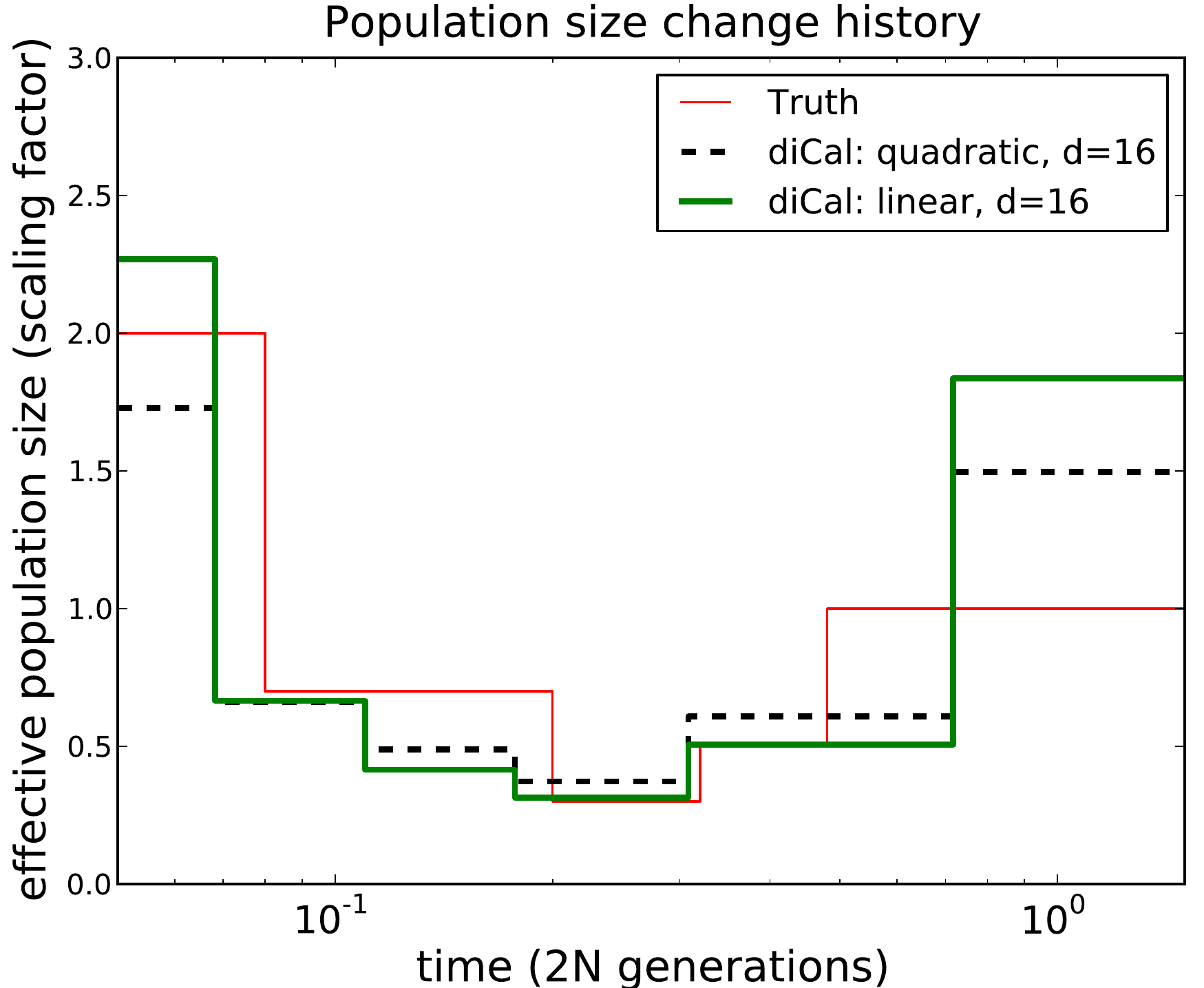}
	\label{fig:results_d16}}
	\caption{Results on simulated data, using the same discretization for the linear and quadratic methods. Each method was run for 20 iterations.
\subref{fig:likelihood} The log likelihood of the EM algorithm, plotted against time, for both the linear and quadratic methods.
\subref{fig:results_d16} Population size change history results for the linear and quadratic methods, run with the same discretization using $d=16$ and estimating $6$ parameters.
	}
	\label{fig:convergence}
\end{figure}

The runtime of the full EM algorithm depends on the convergence of the M-step, which can be variable. 
Occasionally we observed convergence issues for the quadratic method, which requires a multivariate optimization routine.
For the linear method, we used the univariate Brent optimization routine from Apache Math Commons (\url{http://commons.apache.org/proper/commons-math/}), which converges quickly and to a large extent avoids local maxima. 

To examine the convergence of the two EM algorithms, we ran the linear and quadratic methods on the simulated data with $10$ haplotypes and the
same number of intervals $d=16$.
We examine the likelihoods
in Figure~\ref{fig:likelihood}.
The linear method reaches parameter estimates of higher
likelihood, although it is unclear whether the two methods
have found different local maxima, or whether the quadratic method is approaching
the same maximum more slowly.
Figure~\ref{fig:results_d16} shows the inferred population sizes for each method, which although similar, are not identical.

We have also looked at the amount of memory required for each method, and although the difference is small, the linear method does require more memory to store the augmented forward and backward tables. A more thorough investigation of memory requirements will be important as the size of the data continues to increase.

\section{Discussion}
The improvement to diCal described in this paper will enable users to analyze larger datasets and infer more detailed demographic histories. This is especially important given that large datasets are needed to distinguish between histories with subtle or recent differences. By using samples of 10 haplotypes rather than 2, diCal v1.0 \cite{diCal} was able to distinguish between histories that diverged from each other less than $0.1$ coalescent time units ago, in which period PSMC tends to exhibit runaway behavior and hence cannot produce reliable population size estimates. The faster algorithm described here can handle samples of 30 haplotypes with equivalent computing resources. Our results indicate that this improves the method's ability to resolve rapid, recent demographic shifts. 

In organisms where multiple sequenced genomes are not available, the resources freed up by $O(d)$ HMM decoding could be used to avoid grouping sites into 100-locus bins. This binning technique is commonly used to improve the scalability of PSMC, but has the potential to downwardly bias coalescence time estimates in regions that contain more than one SNP per 100~bp.

In general, it is a difficult problem to choose the time discretization that can best achieve the goals of a particular data analysis, achieving high resolution during biologically interesting time periods without overfitting the available data. Sometimes it will be more fruitful to increase the sample size $n$ or sequence length $L$ than to refine the time discretization; an important avenue for future work will be tuning $L,n,$ and $d$ to improve inference in humans and other organisms. 

Another avenue for future work will be to develop augmented HMMs for coalescent models with population structure.  Structure and speciation have been incorporated into several versions of CoalHMM and diCal, and the strategy presented in this paper could be used to speed these up, though a more elaborate network of hidden states will be required. We are hopeful that our new technique will help coalescent HMMs keep pace with the number and diversity of genomes being sequenced and tease apart the demographic patterns that differentiated them.

\vspace{5mm}
\noindent{\bf Acknowledgments.}
 We are grateful to Matthias Steinr\"ucken and other members of the Song group for helpful discussions. This research was supported in part by NSF Graduate Research Fellowships to K.H. and S.S., and by an NIH grant R01-GM094402 and a Packard Fellowship for Science and Engineering to Y.S.S.

\clearpage
\bibliographystyle{splncs03}
\bibliography{refs}

\clearpage 
\appendix
\section{Appendix}    

\subsection{Explicit Computation of Transition Probabilities \label{explicit:transition}}

In Equations 2 and 12 from the main text,
we decompose the transition probabilities $\phi(s \mid s')$
and the stationary probability $\zeta(s)$ into the component terms
$\mathbb{P}(C_i \mid C_{>i-1})$, $\mathbb{P}(C_{>i} \mid C_{>i-1})$,
$\mathbb{P}(R_i ,C_i \mid T_\ell=i),$ $\mathbb{P}(R_i ,C_i \mid T_\ell>i),$ $\mathbb{P}(R_i ,C_{>i} \mid T_\ell=i),$ $\mathbb{P}(R_i ,C_{>i} \mid T_\ell>i)$, and $\mathbb{P}(\overline{R} \mid T_\ell=i)$. Here we give explicit formulae for these transition probabilities in terms of scaling factors $\lambda_1,\ldots,\lambda_d$ that specify the relative effective population sizes within each time interval.  These formulae are specific to the method diCal with variable population size but no population structure. Very similar computations could be used for diCal with population structure, as well as for PSMC, CoalHMM, and related methods.

In addition to $\lambda_1,\ldots,\lambda_d$, these formulae will include the recombination rate $\rho$, scaled with respect to an implicit population size $N_0$ such that $\lambda_i\cdot N_0$ is the effective population size in interval $i$ and $\rho=4N_0r$, where $r$ is the recombination rate per site per generation. Time intervals are defined with respect to a fixed sequence of time points $t_0=0<t_1<\cdots<t_d=\infty$, where the $i$th time state is the interval between $t_{i-1}$ and $t_i$. In addition, $\overline{n}_i$ denotes the average number of lineages that are present at time $t_{i-1}$ in an $n$-leaf coalescent tree, and is computed in \cite{diCal}.

We compute the components of the stationary and transition probabilities as follows:
\allowdisplaybreaks
\begin{eqnarray}
\mathbb{P}(C_{>i}\mid C_{>i-1})
&=& e^{-(t_i-t_{i-1})\overline{n}_i/\lambda_i} \notag \\
\mathbb{P}(C_i \mid C_{>i-1})
&=&1-e^{-(t_{i-1}-t_i)\overline{n}_i/\lambda_i} \notag\\
\mathbb{P}(\overline{R} \mid T_\ell = i) 
&=&\frac{1}{1-e^{-\overline{n}_i(t_i-t_{i-1})/\lambda_i}} \int_{t=t_{i-1}}^{t_i} \frac{\overline{n}_i}{\lambda_i} e^{-\overline{n}_i(t-t_{i-1})/\lambda_i-t\rho}dt \notag \\
&=& \frac{\overline{n}_i}{\overline{n}_i+\lambda_i\rho} \frac{ e^{-t_{i-1}\rho}-e^{-t_i\rho-\overline{n}_i(t_i-t_{i-1})/\lambda_i}}{1-e^{-\overline{n}_i(t_i-t_{i-1})/\lambda_i}} \notag \\
\mathbb{P}(R_i, C_i \mid T_\ell = i)
&=&\frac{1}{\int_{t_{i-1}}^{t_i}\frac{\overline{n}_i}{\lambda_i}e^{-\overline{n}_i (t-t_{i-1})/\lambda_i}dt}\int_{t_0=t_{i-1}}^{t_i}\int_{t=t_{i-1}}^{t_i}\int_{t_r=t_{i-1}}^{t_0\wedge t} \frac{\overline{n}_i}{\lambda_i}e^{-\overline{n}_\ell(t_0-t_{i-1})/\lambda_i} \notag \\
&&\cdot \rho e^{-t_r\rho} \cdot \frac{\overline{n}_i}{\lambda_i} e^{-\overline{n}_i (t-t_r)/\lambda_i}dt_rdtdt_0 \notag \\
&=& \frac{1}{1-e^{-\overline{n}_i(t_i-t_{i-1})/\lambda_i}}\left(\frac{\rho\lambda_i}{\overline{n}_i+\rho\lambda_i} e^{-\rho t_{i-1}}-2e^{-\overline{n}_i(t_i-t_{i-1})/\lambda_i-\rho t_{i-1}}\right. \notag \\
&&\left.-\frac{\rho\lambda_i}{\overline{n}_i-\lambda_i\rho}e^{-\rho t_{i-1}-2\overline{n}_i(t_i-t_{i-1})/\lambda_i}+\frac{2\overline{n}_i^2}{(\overline{n}-\lambda_i\rho)(\overline{n}_i+\lambda_i\rho)}e^{-\rho t_i-\overline{n}_i(t_i-t_{i-1})/\lambda_i}\right) \notag\\
\mathbb{P}(R_i,C_i \mid T_\ell > i)
&=& \int_{t=t_{i-1}}^{t_i} \int_{t_r=t_{i-1}}^{t} \rho e^{-\rho t_r} \cdot \frac{\overline{n}_i}{\lambda_i}e^{-\overline{n}_i(t-t_r)/\lambda_i}dt_rdt \notag \\
&=& e^{-t_{i-1}\rho}+\frac{\lambda_i\rho}{\overline{n}_i-\lambda_i \rho} e^{-\overline{n}_i(t_i-t_{i-1})/\lambda_i-t_{i-1}\rho}-\frac{\overline{n}_i}{\overline{n}_i-\lambda_i\rho}e^{-t_i\rho} \notag \\
\mathbb{P}(R_i, C_{>i} \mid T_\ell = i)
&=&\frac{1}{\int_{t_{i-1}}^{t_i}\frac{\overline{n}_i}{\lambda_i}e^{-\overline{n}_i (t-t_{i-1})/\lambda_i}dt}\int_{t_r=t_{i-1}}^{t_i}\int_{t_0=t_r}^{t_i}\frac{\overline{n}_i}{\lambda_i}e^{-\overline{n}_i(t_0-t_{i-1})/\lambda_i} \notag \\
&&\cdot \rho e^{-t_r\rho} \cdot  e^{-\overline{n}_i (t_i-t_r)/\lambda_i}dt_0dt_r \notag \\
&=&\frac{1}{1-e^{-\overline{n}_i(t_i-t_{i-1})/\lambda_i}}\cdot e^{-\overline{n}_i(t_i-t_{i-1})/\lambda_i}\left(e^{-t_{i-1}\rho}-\frac{\overline{n}_i}{\overline{n}_i-\lambda_i\rho}e^{-t_i\rho}\right.\notag \\
&&\left.+\frac{\rho\lambda_i}{\overline{n}_i-\rho\lambda_i} e^{-\rho t_{i-1}-\overline{n}_i(t_i-t_{i-1})/\lambda_i}\right) \notag
\end{eqnarray}

\begin{eqnarray}
\mathbb{P}(R_i, C_{>i} \mid T_\ell > i)
&=& \mathbb{P}(R_i \mid T_\ell > i) - \mathbb{P}(R_i, C_i \mid T_\ell > i) \notag\\
&=& (1-e^{-(t_i - t_{i-1})\rho}) - (e^{-t_{i-1}\rho}+\frac{\lambda_i\rho}{\overline{n}_i-\lambda_i \rho} e^{-\overline{n}_i(t_i-t_{i-1})/\lambda_i-t_{i-1}\rho}-\frac{\overline{n}_i}{\overline{n}_i-\lambda_i\rho}e^{-t_i\rho}). \notag
\end{eqnarray}

\subsection{Posterior Expectations for the Augmented HMM \label{augmented:posterior:expectations}}

In this section of the appendix, we discuss how to compute the posterior expectations in
\eqref{em:objective:augmented:sub}.
We express these posterior expectations in terms of usual
forward and backward probabilities $f(\subObs{1:\ell}, \state_\ell)$
and $b(\subObs{\ell+1:L} \mid \state_\ell)$,
and also
 the combined forward probabilities
$\feq{i}{\ell}$,  $\fsum{i}{\ell}$, and $\fgreaterthan{i}{\ell}$ introduced in 
Section 2.1 of the main text.
In addition, we need to define the combined backward probabilities,
\begin{align*}
  b(\subObs{\ell+1:L} \mid T_{\ell+1} = i) &= 
  \frac{\mathbb{P}(\subObs{\ell+1:L},
    T_{\ell+1}=i)}{\mathbb{P}(T_{\ell+1} = i)} \\
&=
\frac{f(\subRevObs{1: L-\ell},
    T^{(r)}_{L-\ell} = i)}{\mathbb{P}(T = i)}, \\
  b(\subObs{\ell+1:L} \mid T_{\ell+1} > i) &= 
  \frac{\mathbb{P}(\subObs{\ell+1:L},
    T_{\ell+1}>i)}{\mathbb{P}(T_{\ell+1} > i)} \\
&=\frac{f(\subRevObs{1: L-\ell},
    T^{(r)}_{L-\ell} > i)}{\mathbb{P}(T > i)} \\
  &= \frac{f(\subRevObs{1: L-\ell},
    T^{(r)}_{L-\ell} > i)}{\mathbb{P}(C_{>1}) \mathbb{P}(C_{>2} \mid
  C_{>1}) \cdots \mathbb{P}(C_{>i} \mid C_{>i-1})}.
\end{align*}

We start by showing how to express $\mathbb{E}\Big[ \# \ell:
  \{\recom_{\ell,\ell+1} = R_i, T_\ell>i , T_{\ell+1} > i\}\mid
  \subObs{1:L}  \Big]$ in terms of these forward and backward probabilities:
\begin{align*}
  &  \mathbb{E}\Big[ \# \ell:
  \{\recom_{\ell,\ell+1} = R_i, T_\ell>i , T_{\ell+1} > i\}\mid
  \subObs{1:L}  \Big] \\
  &=
  \sum_{\ell=1}^{L-1}
  \mathbb{P} (\recom_{\ell,\ell+1} = R_i, T_\ell>i , T_{\ell+1} > i\mid
  \subObs{1:L}) \\
  &= \frac1{\pi(\obs)} \sum_{\ell=1}^{L-1} \fsum{i}{\ell}
  \mathbb{P}(R_i, C_{>i} \mid T_\ell > i) \\
  & \qquad \qquad \times
  \sum_{j > i} \left(\prod_{k=i+1}^{j-1} \mathbb{P}(C_{>k} \mid
    C_{>k-1}) \right)
  \mathbb{P}(C_j \mid C_{>j-1}) \bbP(\subObs{\ell+1:L} \mid T_{\ell+1} = j)
  \\
  &= \frac1{\pi(\obs)} \sum_{\ell=1}^{L-1} \fsum{i}{\ell}
  \mathbb{P}(R_i, C_{>i} \mid T_\ell > i)
  \sum_{j>i} \frac{\mathbb{P}(\subObs{\ell+1:L}, T_{\ell+1} =
    j)}{\mathbb{P}(C_{>1}) \mathbb{P}(C_{>2} \mid C_{>1}) \cdots
    \mathbb{P}(C_{>i} \mid C_{>i-1})}
\\
  &= \frac1{\pi(\obs)} \sum_{\ell=1}^{L-1} \fsum{i}{\ell} 
  \mathbb{P}(R_i, C_{>i} \mid T_\ell > i)
  b(\subObs{\ell+1 : L} \mid T_{\ell+1} > i),
\end{align*}
where $\pi(\obs) := \mathbb{P}(\obs) = \sum_s f(\subObs{1:L},s)$.

Computing the other posterior expectations is similarly straightforward.
We list the derived expressions for them here:
\begin{align*}
    & \mathbb{E} \Big[ \# \ell: \{\recom_{\ell,\ell+1} = \overline{R}, T_\ell = i\} \mid
  \subObs{1:L} \Big] 
\\ & \qquad \qquad \qquad =  \frac1{\pi(x)} \sum_{\ell=1}^{L-1} \sum_{h
    \in \mathcal{H}} f(\subObs{1:\ell}, (h,i))
    \bbP (\overline{R} \mid T_\ell = i) \xi(\subObs{\ell+1} \mid \sub{h}{\ell+1},i)
    b(\subObs{\ell+2:L} \mid (h,i)) ,
\\
  &\mathbb{E}\Big[ \# \ell:
  \{\recom_{\ell,\ell+1} = R_i, T_\ell = T_{\ell+1} = i\} \mid
  \subObs{1:L}  \Big] 
\\ & \qquad \qquad \qquad = 
\frac1{\pi(x)} \sum_{\ell=1}^{L-1} f(\subObs{1:\ell}, T_\ell=i) \mathbb{P}(R_i,C_i \mid T_\ell= i)
 b(\subObs{\ell+1:L} \mid T_{\ell+1}=i),
\\
  &\mathbb{E}\Big[ \# \ell:
  \{ \recom_{\ell,\ell+1} = R_i, T_\ell > T_{\ell+1} = i \} \mid
  \subObs{1:L}  \Big] 
\\ & \qquad \qquad \qquad = 
\frac1{\pi(x)} \sum_{\ell=1}^{L-1} f(\subObs{1:\ell}, T_\ell > i) \mathbb{P}(R_i,C_i \mid
T_\ell>i) 
 b(\subObs{\ell+1:L} \mid T_{\ell+1} = i),
\\
  &\mathbb{E}\Big[ \# \ell:
  \{\recom_{\ell,\ell+1} = R_i, T_{\ell+1} > T_{\ell} = i\}\mid
  \subObs{1:L}  \Big]
\\ & \qquad \qquad \qquad =
\frac1{\pi(x)} \sum_{\ell=1}^{L-1} f(\subObs{1:\ell}, T_\ell=i) \mathbb{P}(R_i, C_{>i}
\mid T_\ell = i) 
b(\subObs{\ell+1:L} \mid T_{\ell+1} > i),
\\
  &\mathbb{E}\Big[ \# \ell:
  \{\recom_{\ell,\ell+1} = R_i, T_\ell>i , T_{\ell+1} > i\}\mid
  \subObs{1:L}  \Big]
  \\ & \qquad \qquad \qquad = 
\frac1{\pi(x)} \sum_{\ell=1}^{L-1} f(\subObs{1:\ell}, T_\ell > i)
  \mathbb{P}(R_i, C_{>i} \mid T_\ell > i)
  b(\subObs{\ell+1 : L} \mid T_{\ell+1} > i),
\\
  &\mathbb{E}\Big[ \# \ell:
  \{\recom_{\ell,\ell+1} \in R_{<i}, T_{\ell+1} > i\}\mid \subObs{1:L}
  \Big]
\\ & \qquad \qquad \qquad = 
\frac1{\pi(x)} \sum_{\ell=1}^{L-1} 
\fgreaterthan{i-1}{\ell}
\mathbb{P}(C_{>i} \mid
C_{>i-1})
b(\subObs{\ell+1:L} \mid T_{\ell+1} > i),
\\
  &\mathbb{E}\Big[ \# \ell:
  \{\recom_{\ell,\ell+1} \in R_{<i}, T_{\ell+1} = i\}\mid \subObs{1:L} 
   \Big] 
\\ & \qquad \qquad \qquad =
\frac1{\pi(x)} \sum_{\ell=1}^{L-1} 
\fgreaterthan{i-1}{\ell}
\mathbb{P}(C_{i} \mid
C_{>i-1})
b(\subObs{\ell+1:L} \mid T_{\ell+1} = i),
\\
& \mathbb{P}(T_1 = i \mid \subObs{1:L}) 
\\ & \qquad \qquad \qquad =
\frac1{\pi(x)} \sum_{h \in \mathcal{H}} f(\subObs{1}, (h,i))
b(\subObs{2:L} \mid (h,i)),
\\
&\mathbb{P}(T_1 > i \mid \subObs{1:L}) 
\\ & \qquad \qquad \qquad =
\frac1{\pi(x)} \sum_{h \in \mathcal{H}} \sum_{j=i+1}^d f(\subObs{1},
(h,j)) b(\subObs{2:L} \mid (h,j))
\\ & \qquad \qquad \qquad = 
\mathbb{P}(T_1 > i+1 \mid \subObs{1:L}) + \mathbb{P}(T_1 = i+1 \mid \subObs{1:L}).
\end{align*}

\subsection{Running diCal on Simulated and Real Data}

%

To run diCal on the simulated data, we used the commandlines below ($d=9$ and $d=21$):
\begin{verbatim}
java -jar diCal.jar -F data.fasta -I params.txt -n 10 -p "3+3+3" -t 1.0 -u 0
java -jar diCal.jar -F data.fasta -I params.txt -n 10 -p "8+2+2+2+2+2+3" -t 2.0 -u 1
\end{verbatim}

\noindent where $n$ is the total number of haplotypes, $p$ is the grouping of discretization intervals into parameters, $t$ is the start point of the last discretization interval (in coalescent units) and $u$ is a flag for using the linear vs. quadratic method. \\

To run diCal on the 1000 Genomes data, we used the commandlines:
\begin{verbatim}
java -jar diCal.jar -F data.fasta -I params.txt -n 10 -p "3+3+3" -t 1.0 -u 0
java -jar diCal.jar -F data.fasta -I params.txt -n 10 -p "5+3+2+2+2+3+4" -t 2.0 -u 1
\end{verbatim}

\noindent For both the simulated and real data, the parameter groupings were chosen such that the number of parameters inferred would be $d/3$, with minimal runaway behavior.

\end{document}